\documentclass[twocolumn,showpacs,preprintnumbers,amsmath,amssymb]{revtex4}

\usepackage{graphicx}
\usepackage{dcolumn}
\usepackage{bm}

\begin{document}

\title{Nonlinear optical responses and gap-soliton lattices in layered dielectric materials.}

\author{Alain M.~Dikand\'e}
\affiliation{D\'epartement de Physique, Facult\'e des Sciences, Universit\'e de Douala BP 24157 Douala, Cameroun.}    
 
 \author{ Timol\'eon C. Kofan\'e}
\affiliation{Laboratoire de M\'ecanique, D\'epartement de Physique, Facult\'e des sciences, Universit\'e de yaound\'e BP 812 Yaound\'e, Cameroun.}%
\date{\today}

\begin{abstract}
A new approach to the nonlinear reponse of a multiplayer structure consisting of alternating dielectric materials with intensity-dependent 
dielectric constants is considered in both analytical and numerical viewpoints. The nonlinear feedback of the structure gives rise to 
an artificial potential acting as a mean-field feedback on each single dielectric material due to other layers. For superlattices 
resulting from a regular periodic arrangement of an infinite number of single dielectric layers, it is shown that an analytical 
expression of this mean-field feedback can be obtained and waveshapes of the associate gap solitons are derived. The case of a multilayer 
with dispersive interlayer interactions is considered numerically and leads to strongly localized gap solitons. Implicatons 
of the artificial soliton structure as for possible theoretical modelling of soliton-compression 
phenomena are discussed. 
\end{abstract}

\pacs{42.55.Tv, 42.65.Sf, 42.65.Tg}

\maketitle
Superlattices with Bragg reflectors are generally not accessible to any light frequency. In multilayer structures 
consisting of a periodic arrangement of single dielectric materials, the dependence of dielectric constants on 
the light intensity affects the intimate structure of materials leading to the formation of defects which are reponsible 
for these apparent gap excitations. Chen and Mills~\cite{chen} pointed out a mechanism so-called stop-gap radiation 
transmission following 
which the light could still propagate within the superlattice due to partial closing of the gap defect. Later, it was 
suggested\cite{qiming,lidoriskis} that a suitable arrangement of single dielectric materials 
composing a superlattice structure too may favor gap-soliton propagation. In recent years, a great deal of interest has been devoted to understanding 
nonlinear features of photonic band-gaps~\cite{yablo,john,joan} to both view points of their structural and dynamical 
properties. Crudely speaking, 
a nonlinear photonic band-gap excitation is an electromagnetic wave moving trapped to a defect in the dielectric 
material with an exponentially localized shape. \\ 
Our aim in this work is to suggest a new approach to the global nonlinear feedback resulting from the superposition of 
the optical nonlinear properties of each single dielectric layer in a multilayer structure. In general, the propagation of 
electromagnetic field in a single dielectric layer is governed by the nonlinear Schr\"odinger equation(NLSE) i.e.~\cite{chen}:
\begin{equation}
i\phi_t + \phi_{zz} + J_o\mid \phi \mid^2 \phi=0. \label{eqat1}
\end{equation}
 Solutions of this equation which take account of the 
localized feature of the light due to a local nonlinear feedback are stationary solitons. For a self-focusing 
nonlinear feedback, the stationary soliton is bell-shaped(pulse) and is obtained as:
\begin{eqnarray}
\phi(z,t)= u(z) \, exp(-i \omega t),   \nonumber \\
u(z)= u_o \, sech (\sqrt{\omega} \, z ), \hspace{.2in} u_o^2= \frac{2 \omega}{J_o}, \label{founde1} 
\end{eqnarray}
  
\begin{figure}
\includegraphics[height=5cm]{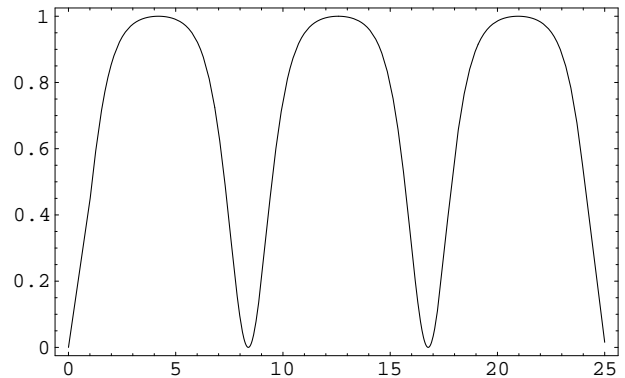}
\caption{\label{fig:pot} The global nonlinear feedback potential. }
\end{figure}

\begin{figure}
\includegraphics[height=5cm]{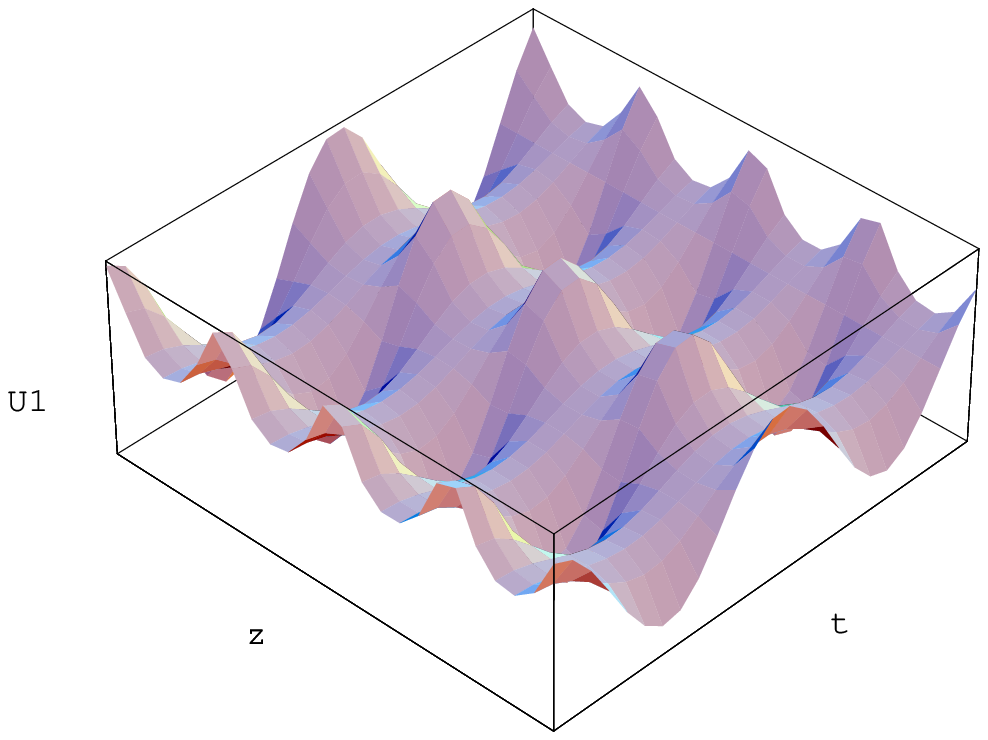}
\caption{\label{fig:solit2a} Spatio-temporal waveshape of the $dn$ gap-soliton lattice. }
\end{figure}

\begin{figure}
\includegraphics[height=5cm]{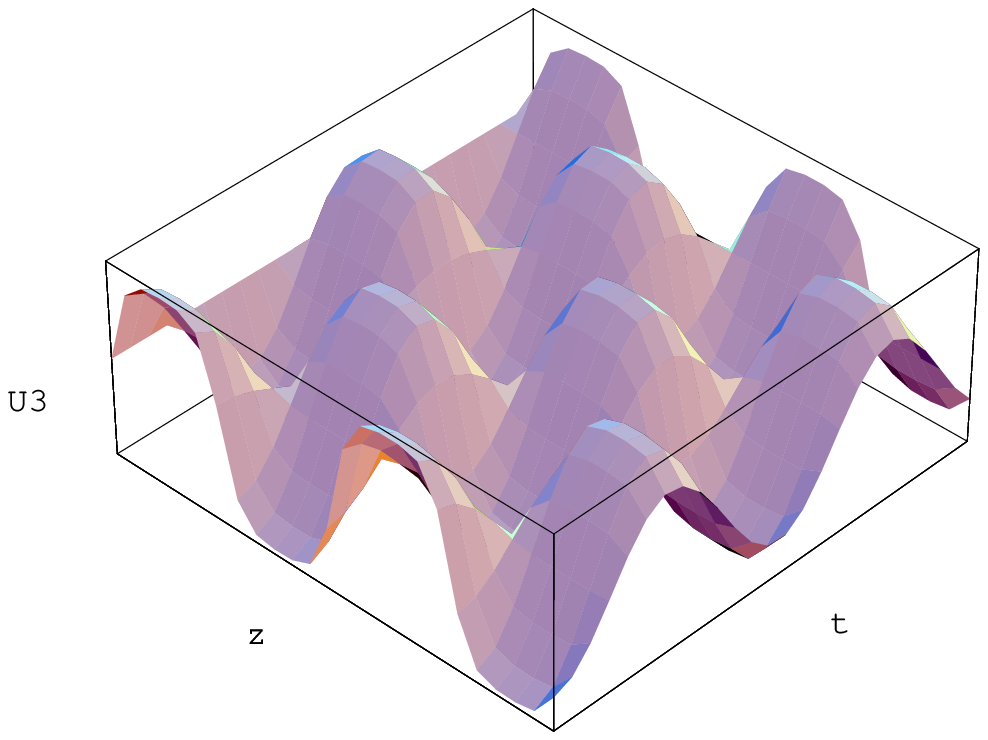}
\caption{\label{fig:solit2b} Spatio-temporal waveshape of the $sn$ gap-soliton lattice. }
\end{figure}

\begin{figure}
\includegraphics[height=5cm]{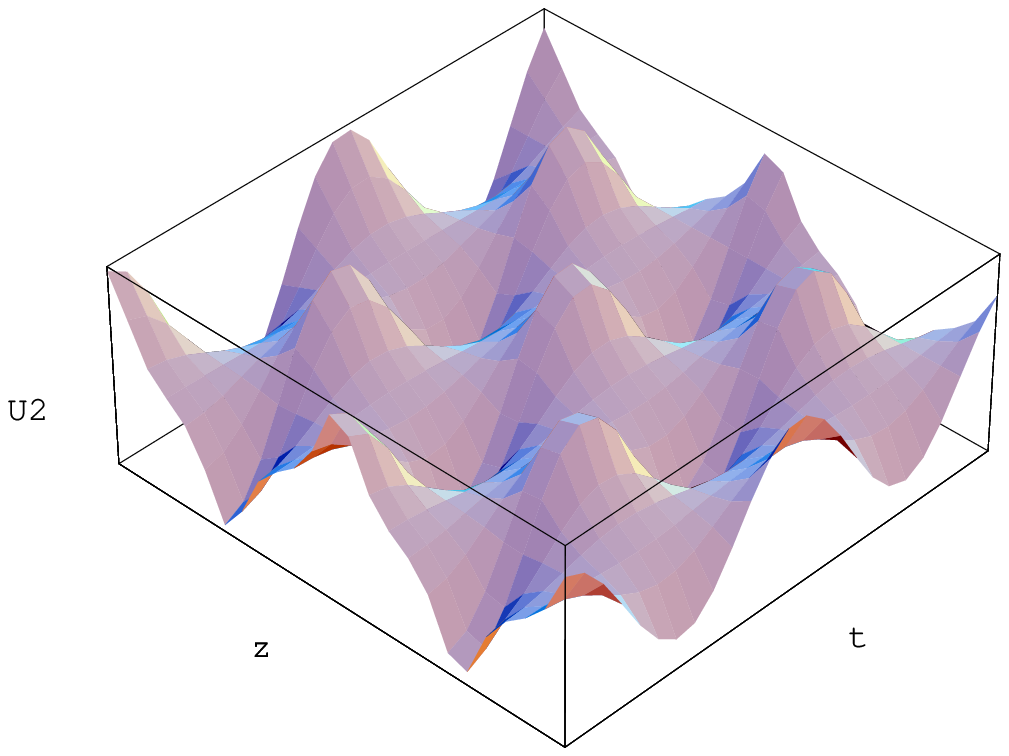}
\caption{\label{fig:solit2c} Spatio-temporal waveshape of the $cn$ gap-soliton lattice. }
\end{figure}

In the spirit of the NLSE, this solution describes an harmonic time oscillation hidden in a spatial shell with a 
nonlinear topological shape. If we assume $\omega$ to carry the optical information, then by virtue of its topological 
feature the nonlinear spatial excitation will act as a shell protector. To a mathematical viewpoint,~(\ref{founde1}) 
is a spatial topological soliton with an internal frequency hence a breather soliton. Refering to the NLSE and from the
 analytical expression of the soliton waveshape, $\omega$ will depend on the intrinsic parameters 
of the breather. Thus~(\ref{founde1}) is a gap soliton, where $u_o$ is the amplitude of the soliton gap. \\
If we now assume a superlattice resulting from periodic arrangement of single dielectric layers obeying 
equation~(\ref{eqat1}), gap solitons of each of these single dielectric materials become their foundamental modes. 
However, owing to the broken translational invariance from the coupling of multiple layers, these foundamental modes will 
no more obey single NLSEs and the dynamics of a given $n^{th}$ optical component is from now on governed by the following equation:
\begin{equation}
i\phi_{n t} + \phi_{n zz} + \sum_{m=n}^{N}{J_{m-n}\mid \phi_m \mid^2 \phi_n}=0. \label{eqat2}
\end{equation}
We will assume all dielectric materials to be identical and denote their common foundamental frequency by $\Omega_o$. The 
key point of our study 
is the assumption that equation~(\ref{eqat2}) is self-consistent, i.e. is valid for all optical components. Namely, 
the limit $N=m$ should reproduce the single-pulse soliton~(\ref{founde1}). In this context, the sum in $m$ 
runs over the entire superlattice 
and therefore takes into account all foundamental modes. Doing so, the problem turns to that of a single dielectric layer 
in a potential bath(hereafter denoted "mean-field potential") created by its surroundings. This 
mean-field potential is adiabatic in the sense that it is switched on after soliton defects have formed on all the $N$ dielectric 
layers. To allow for the interaction among solitons, we need to separate their centres of 
mass from their spatial coordinates on layers. In fact, this amounts to introduce a pinning-center in the overall 
coordinate of each foundamental soliton so that the isolated single-pulse is readily rewritten:
\begin{equation}
u(z)= u_o \, sech \sqrt{\Omega_o}(z - z_n) . \label{founde2} 
\end{equation}
With this more appropriate shape, equation~(\ref{eqat2}) leads to the following equation for the spatial wavefunction of 
the light wave passing through a given $n^{th}$ optical component and interacting with an $m^{th}$ one via a coupling $J_{m-n}$:  
\begin{equation}
u_{n zz} + \left[ \omega_n + \frac{\Omega_o}{J_o} \sum_{m=n}^{N}{J_{m-n} sech^2 \sqrt{\Omega_o}(z - z_m)} \right] u_n=0. \label{eqat3}
\end{equation}
Below, nonlinear solutions of this equation are discussed assuming two different contexts. We first consider the case of a 
uniform periodic arrangement of dielectric layers(regularly distributed single-layer feedbacks), and next the case of a dispersive 
coupling(dispersion-induced localized single-layer feedbacks) between optical components. \\
For $J_{m-n}=J_o$ where $J_o$ is a constant, the mean-field potential experienced by the $n^{th}$ optical component 
is the sum over $sech^2$ functions. This sum is exact if we note the following relevant identity($\ell=m - n$):
\begin{widetext}
\begin{eqnarray}
\sum_{\ell=-\infty}^{\infty}{sech^2\left[\sqrt{\Omega_o}(z - \ell L)\right]} = \left(\frac{2}{\pi}\right)^2E'K' - \left(\frac{\nu K'}{E'}\right)^2sn^2\left(\frac{z}{d_{\nu}} \right), \nonumber 
\\
 d_{\nu}= \frac{\pi}{2K' \sqrt{\Omega_o}}, \hspace{.2in} L= 2K d_{\nu}, \hspace{.2in} 0 \leq \nu \leq 1.           \label{som2}
\end{eqnarray} 
\end{widetext}
where $K$, $K'$ and $E'$  are Complete Elliptic Integrals and $sn$ is the Jacobi elliptic function of modulus 
$\nu$ and period $L$. This relation follows from simple transformations of the usual Fourier series 
representation of Jacobi Elliptic functions~\cite{abram}, taking into consideration the extra argument $z_n$. 
The above identity, given in terms of the intensities of the local foundamental light waves, 
mimics a periodic lattice of single pulses mutually trapped at their centre-of-mass $z_n= n L$ where 
$L$ is the separation between two interacting pulses. $L$ increases as $\nu$ is increased so that one can monitor the 
degree of compression of single pulses within the potential, namely by tuning the thickness $d_{\nu}$ of the 
mean-field potential wells which according to~(\ref{som2}) is proportional to the width of the single-pulse 
defect(${\frac{1}{\sqrt{\Omega_o}}}$). On fig.~\ref{fig:pot} we sketch the space-dependent part of the global feedback 
potential(i.e. the $sn^2$ function). When $\nu=1$ the mean-field potential reduces to $sech^2 \sqrt{\Omega_o} z$. In connection 
with this finding, below we will also see that we recover the original single-pulse soliton. Remark that in this limit 
the separation $L$ between the single pulses is infinite. \\
Replacing the identity~(\ref{som2}) in equation~(\ref{eqat3}) we find:
\begin{eqnarray}
\left[\frac{\partial^2}{\partial x^2} + P_{\nu}(\omega_n) - 2\nu^2 \, sn^2(x) \right] u_n(x)= 0, \nonumber 
\\
P_{\nu}(\omega_n)= \left(\frac{\pi}{2K'}\right)^2 \frac{\omega_n}{\Omega_o} + \frac{2E'}{K'}, \hspace{.1in} x= \frac{z}{d_{\nu}}, \label{lam}
\end{eqnarray}
 Equation~(\ref{lam}), so-called first-order Lam\'e equation~\cite{lame}, is the actual one governing the spatial 
waveshape of the bunch of individual solitons passing through the $n^{th}$ optical component of the periodic lattice. This 
equation possesses three distinct bound states listed as~\cite{sutherland,dika1,dika2}; 
\begin{eqnarray}
u_1(x)&=& u_{o1}(\nu) \, dn(x), \nonumber 
\\
\omega_o&=& \Omega_oK' \left[k^2K' - 2E' \right] \left(\frac{2}{\pi}\right)^2 ,  \label{soluto}
\end{eqnarray}
\begin{eqnarray}
u_2(x)&=& u_{o2}(\nu) \, sn(x), \nonumber 
\\
\omega_1&=& \Omega_oK' \left[(1+k^2)K' - 2E' \right] \left(\frac{2}{\pi}\right)^2.   \label{solut1}
\end{eqnarray}
\begin{eqnarray}
u_3(x)&=& u_{o3}(\nu) \, cn(x), \nonumber   
\\
\omega_2&=& \Omega_oK' \left[K' - 2E' \right] \left(\frac{2}{\pi}\right)^2,     \label{solut2}
\end{eqnarray}
In general, these bound states form a complete orthonormal set and are normalized accordingly. However, to avoid a divergent 
light intensity we bind their amplitudes to the amplitude of the single-soliton gap $u_o$. The normalization 
condition then becomes:
\begin{equation}
\int_{-L'/2}^{L'/2}{u_i(x) \, u_j(x) dx}= u_o^2 \delta_{i, j}, \hspace{.1in}  L' = \frac{L}{d_{\nu}}, \label{norme}
\end{equation}
and leads to: 
\begin{eqnarray}
u_{o1}^2 &=& \frac{u_o^2}{E},  \nonumber   
\\
u_{o2}^2 &=& \frac{u_o^2 \nu^2}{\left[K - E \right]}, \nonumber
\\
u_{o3}^2 &=& \frac{u_o^2\nu^2}{\left[E - (1 - \nu^2)K \right]}, \label{amplit}   
\end{eqnarray}
We can check that in the limit $\nu \rightarrow 1$ these bound states reduce to:
\begin{eqnarray}
u_1(z)&=& u_3(z)= u_o \, sech \sqrt{\Omega_o} z , \hspace{.1in} \omega_o= \omega_2= -\Omega_o, \nonumber 
\\
u_2(z)&=& 0, \hspace{.2in} \omega_1= 0. \label{asympt}
\end{eqnarray}
On figs.~\ref{fig:solit2a}, ~\ref{fig:solit2b} and~\ref{fig:solit2c} we display the spatio-temporal waveshapes($\nu=0.98$) of 
the three bound states. 


\begin{figure}
\includegraphics[height=5cm]{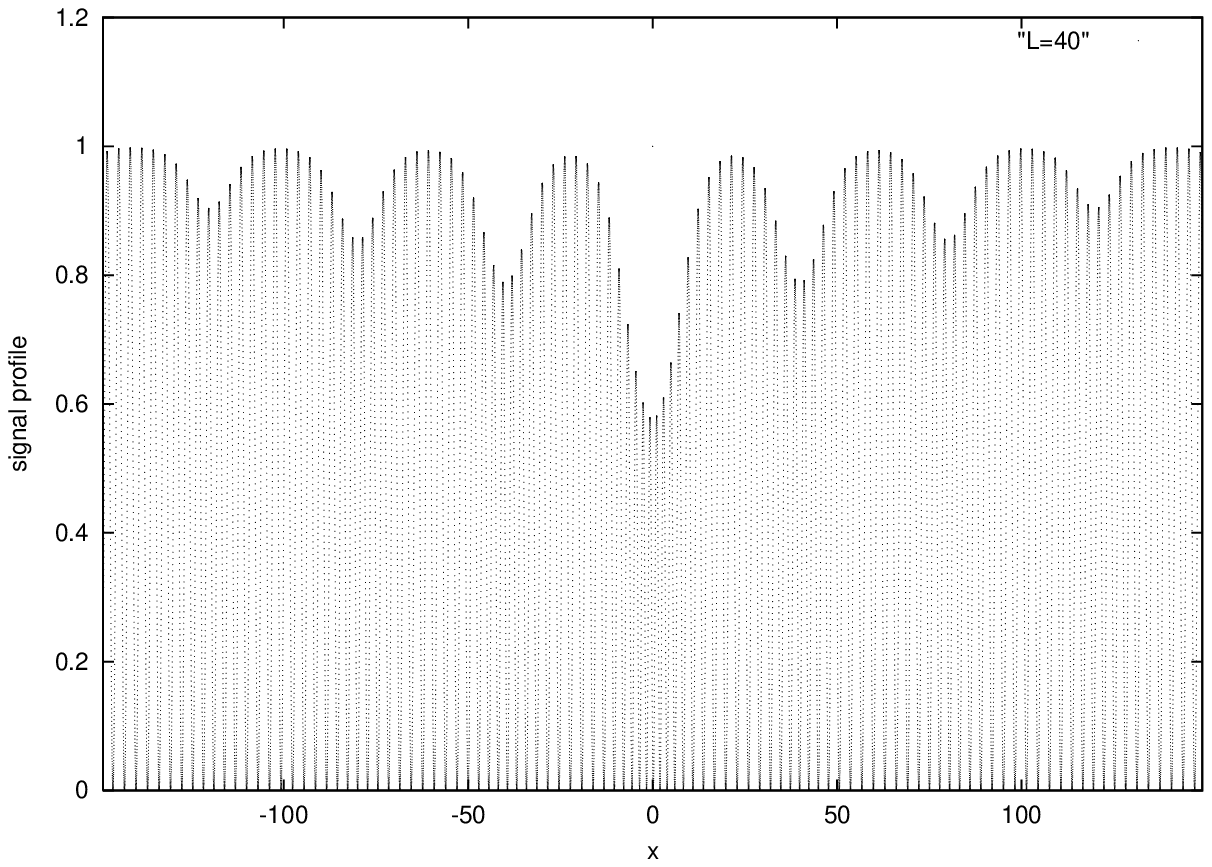}
\caption{\label{fig:solita} Interlayer-dispersion-induced localization of the gap-soliton lattice: $L=40$.}
\end{figure}

\begin{figure}
\includegraphics[height=5cm]{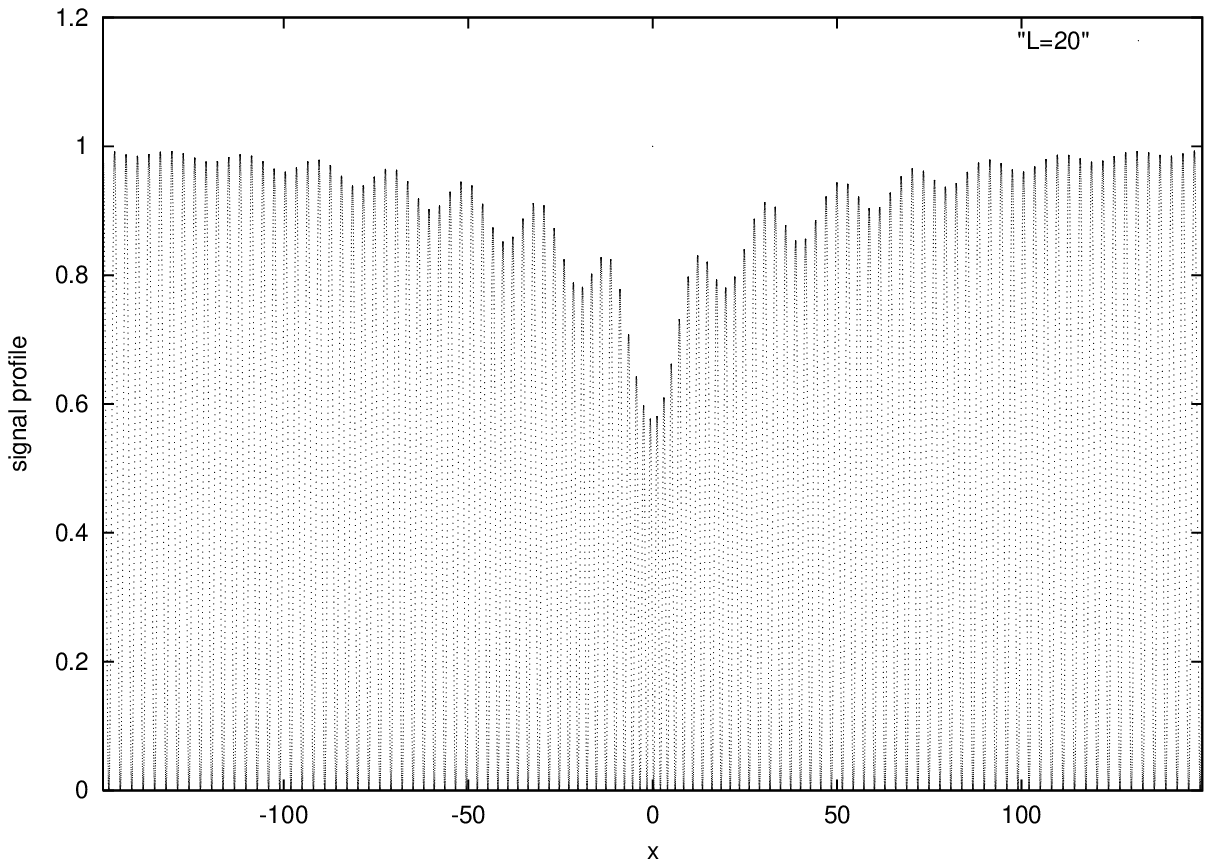}
\caption{\label{fig:solitb} The gap-soliton lattice for $L=20$.}
\end{figure}

\begin{figure}
\includegraphics[height=5cm]{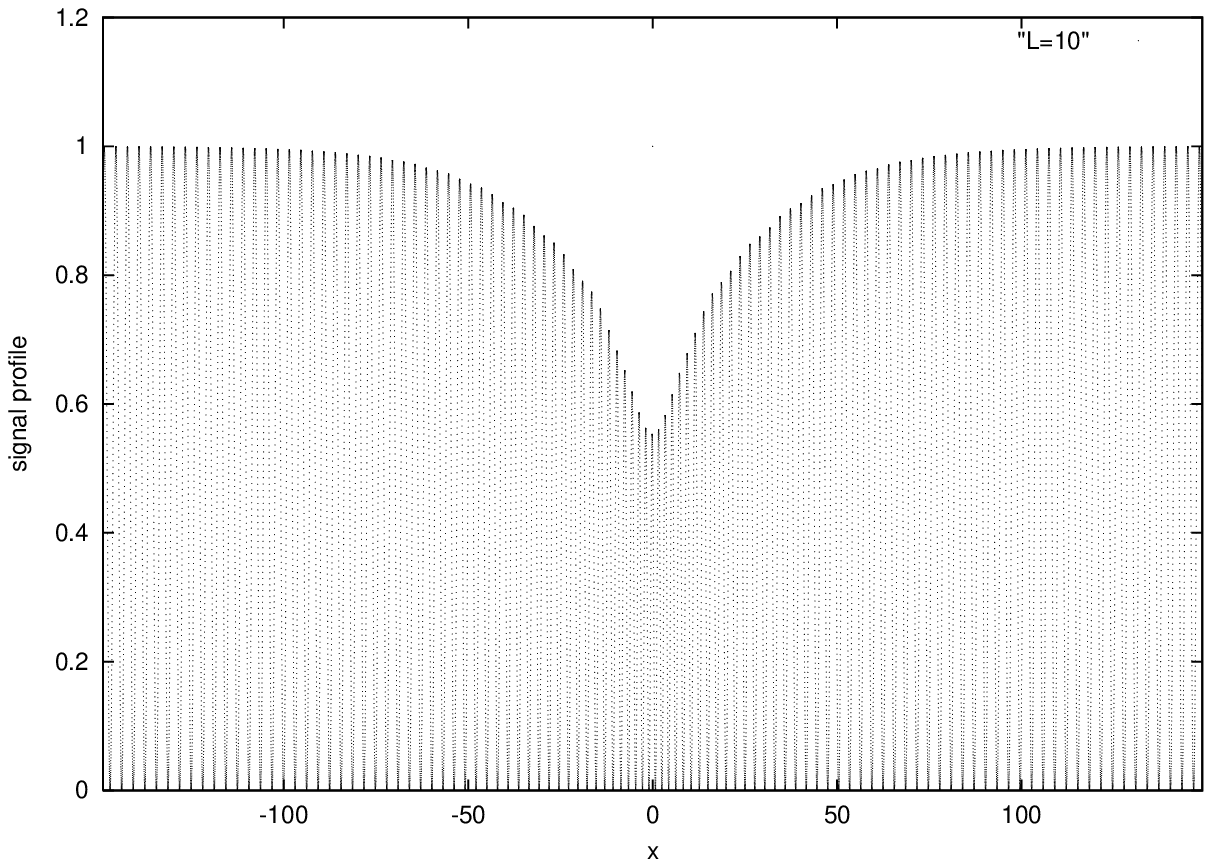}
\caption{\label{fig:solitc} The gap-soliton lattice for $L=10$.}
\end{figure}

\begin{figure}
\includegraphics[height=5cm]{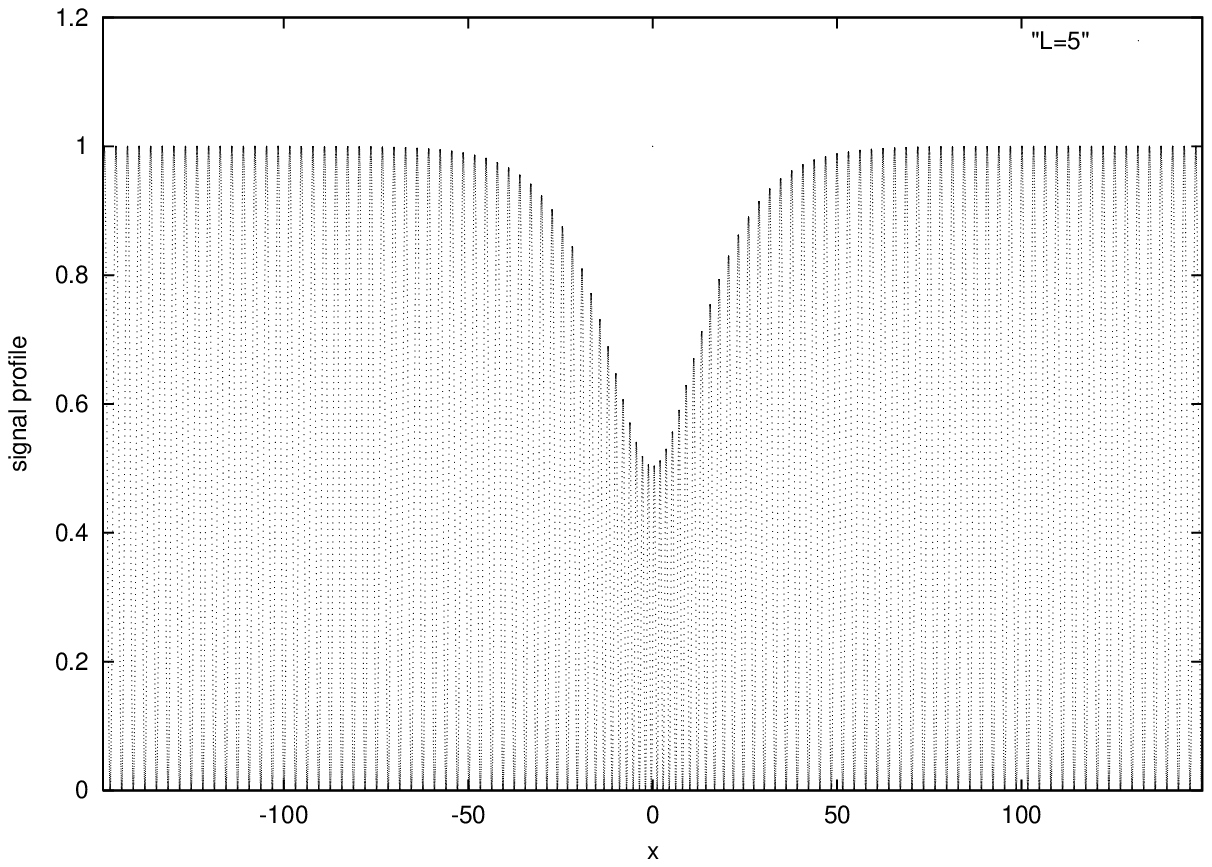}
\caption{\label{fig:solitd} The gap-soliton lattice for $L=5$.}
\end{figure}

According to their asymptotic behaviours, the first and third bound states are two degenerate periodic gap solitons. The 
second bound state, which gives rise to a vanishing single kink in the limit $L \rightarrow \infty$, is the state of 
broken translational symmetries of individual solitons due to the correlations of their (non-zero)centre-of-mass 
coordinates. The vanishing of this mode in the limit $\nu \rightarrow 1$(where the separation between single pulses 
becomes infinite) is a clear evidence of restored translational invariance. \\
 Let us now turn to the case of a dispersive coupling. We will assume the simplest case of an exponentially decreasing interaction 
between single dielectric layers i.e. $J_{m-n}= J_o e^{-\lambda \vert m - n \vert}$~\cite{bishop1}. 
Since the problem cannot be efficiently treated by anlytical means, we followed a numerical approach. 
Figs.~\ref{fig:solita}, ~\ref{fig:solitb}, ~\ref{fig:solitc} and~\ref{fig:solitd} are shapes of 
the light intensity on a single layer within the multilayer structure, simulated assuming four different values of the separation 
$L$ between $N=100$ single pulses. 
The pulse width is 10 units correponding to a foundamental frequency $\Omega_o \sim 0.01$, while the soliton-lattice frequency 
$\omega_n = 10 \Omega_o$. $\lambda$ is set to $0.5$ which we expect to ensure relatively long distance effects.
 The most stricking feature in these figures is a less and less periodic 
distribution of single pulses in the gap-soliton lattice as their separation $L$ decreases. The gap-soliton lattice is clearly seen to transform 
into a bunched soliton state showing strong localization on single layers as solitons in the bunch get more and more compressed. 
Remark that fig.~\ref{solitc} 
corresponds to an inter soliton separation which is equal the single-pulse width. Actually, the behaviours emerging from the figures
 are not surprising since it is a natural fact that a decrease of $L$ will enhance sensibility of the soliton shape to dispersive 
 interlayer interactions.  \\
As a concluding remark, the new analytical gap-soliton derived above are artificial objects consisting of a bunch of single 
localized gap-solitons which are compeled to share the same propagation medium, as opposed to the well-known periodons of 
finite single dielectric layers see e.g. Lidoriskis et al.~\cite{lidoriskis}. Therefore, they can be looked on as 
compressed single gap-solitons where the global mean-field feedback plays role of a compression potential. Since this global 
feedback forms from foundamental modes of each single dielectric layer, the thickness of its wells and amplitude of the 
potential barriers will be fixed by the single-soliton width and gap, respectively. That is to say, one can monitor the 
soliton compression within the global feedback. According to our analytical soliton solutions, the waveshape acquired by the gap soliton 
passing through each dielectric layer reflects features of the compression potential, thus each of these layers 
actually convey a bunch of compressed single pulses. Within the bunch state the soliton gap is lowered, 
decreasing with a decrease of the modulus $\nu$. Recall that a decrease of $\nu$ brings single solitons closer and closer. 
As the dispersive inter-layer interaction just sharpens the soliton-lattice on 
single layers, we also expect similar features(namely a decrease of the soliton gap) to show up in this last context for 
a relatively large number of interacting single dielectric layers.            
\begin{acknowledgments}
 Part of this work was carried out as A. M. Dikand\'e was guest at the Abdus Salam International 
Centre for Theoretical Physics(ASICTP) Trieste, Italy.
\end{acknowledgments}
 
\end{document}